\documentclass[aps,prl,showpacs,showkeys,twocolumn]{revtex4-1} 
\usepackage{graphicx,amsmath,hyperref}
\usepackage{epstopdf}

\newcommand{\beq}{\begin{equation}}
\newcommand{\eeq}{\end{equation}}
\newcommand{\ba}{\begin{eqnarray}}
\newcommand{\ea}{\end{eqnarray}}
\newcommand{\bea}{\begin{eqnarray}}
\newcommand{\eea}{\end{eqnarray}}
\newcommand{\bma}{\begin{subequations}}
\newcommand{\ema}{\end{subequations}}
\newcommand{\bwt}{\begin{widetext}}
\newcommand{\ewt}{\end{widetext}}
\def\abs#1{|\,#1\,|}

\begin{document}

\title{An analytical solution for light field modes in waveguides with nonideal cladding}

\author{N.M.Arslanov$^{1}$,  Ali A.Kamli$^{2}$, and S.A.Moiseev$^{1,3}$}
\email{samoi@yandex.ru}
\affiliation{$^{1}$Kazan Quantum Center, Kazan National Research Technical University, 10 K. Marx, Kazan, 420111, Russia}
\affiliation{$^{2}$Department of Physics, Jazan University, Jazan 28824 P O Box 144 Saudi Arabia}
\affiliation{$^{3}$Zavoisky Physical-Technical Institute of the Russian Academy of Sciences, Kazan, Russia}

\date{\today}


\begin{abstract}

We have obtained an analytical solution
for the dispersion relation of the light field modes
 in the nanowaveguide structure.  The solution has been analyzed for the planar waveguide with metamaterial claddings and dielectric core.
 The analytical solution is valid within the broadband spectral range and is confirmed by existing numerical calculations.
 The developed theoretical approach opens vast possibilities for the analytical investigations of the light fields in the various waveguides.
\end{abstract}

\maketitle

\textit{Introduction.} Control,  manipulation and  spatial confinement of  light fields are the main objectives of photonics where various waveguides and resonator structures play a major role \cite{Hunsperger2009, Tanzilli2012, Haroche2006}.
The growing interest in this field of research is spurred by the invention of diverse nanowaveguide systems in optics, photonics, plasmonics and its applications to manipulate the single photon fields   
\cite{Segal2015, PhysRevLett.112.167401, PhysRevLett.111.247401, PhysRevLett.111.046802, PhysRevLett.111.090502, Tame2013}.
Great expectations are associated here with using new materials, such as photonic crystals, metamaterials, graphene, and others 
\cite{Veselago1967, PhysRevLett.85.3966, PhysRevLett.112.137401, Tame2013, PhysRevLett.111.247401, Maradudin2011, PhysRevLett.113.037401, PhysRevX.4.041042, Richardson2014} causing the detection of entirely new properties of light. 
The strong stream of these experimental investigations stimulates elaborating  the new  theoretical approaches  which could provide a clearer physical picture and  high mathematical precision  for the description of the light fields in the new and well-known  waveguides. 

 Theoretical study of the  light field localized  in the waveguide structures is a long-term theoretical problem \cite{Rayleigh1897, Sommerfeld1909, Stratton1941, Kogelnik1988, Yeh2008, Hunsperger2009, 1132809}. 
Work on the spatial confinement of the free light fields  into the waveguide with the finite cross-section shows a dramatic change in the fundamental properties of light. 
The transversely confined photon wave packet propagating along $z-$direction   acquires the spectral dispersion 
in the  form of  Hamiltonian function
  $\hbar \omega=\sqrt{\left( \hbar ck_{z} \right)^2+W_{n}}$
   corresponding to the relativistic particle with a finite mass $m_0$ \cite{DeBroglie1941, Rivlin1997, Sambles2015}, where $W_{n}=(m_o c^2)^2=\left(\hbar ck_{n}^{\perp}\right)^2$,  $\omega=2\pi c/\lambda$, $c$ is a speed of light, $\lambda$ its wavelength in the vacuum,  $\hbar$ - Planck constant, $k_{z}$-longitudinal wave-number of the $n$-th light mode, and  
$k_{n}^{\perp}$  is a transverse wave number determined by the specific properties of the waveguide (see later). 
It is worth to note the recent experiment \cite{Klaers2010} where effective nonzero photon mass revealed itself in a possibility of Bose-Einstein condensation of photons in the optical microcavity.

In practice, the waveguides have various geometries, cores and materials of the waveguide claddings  \cite{Yeh2008} which are characterized by different spectral properties of permittivity $\epsilon(\omega)$ and magnetic permeability $\mu(\omega)$.
Geometrical and physical properties of the waveguides determine highly specific behavior of the light modes $\psi_{n}$ at the
interface between the claddings and cores identifying the effective cross-section of the confined light modes, the concrete  mass $m_o$  of photon and its minimum frequency $ \omega_{n}$, respectively. 
In particularly, $\omega_{n}=\frac{\pi n}{L} c$ and   
$m_{0}=\frac{\pi n \hbar}{L c}$ for the simplest planar waveguide with the transverse width $L$ and a perfect metal cladding. 

The confined light modes  in the arbitrary waveguides are described by the transcendental dispersion equations \cite{Sommerfeld1909, Yeh2008} that do not have analytical solutions even in the simple cases.
Unfortunately, the transcendental equation is also impracticable for intuitive analysis of the light properties. 
In turn,  the understanding of the eigen mode dispersion relation is required  for the further study of the principal information about the localized electromagnetic fields such as the phase and group velocities, transverse shape and propagation length of the low-losses modes, in particular for the planar or nanowire arrays and  for the  nanoplasmonic, V-groove structures, etc. (see \cite{PhysRevLett.111.046802} \cite{Berini2009} and references there). 

\begin{figure}
\centerline{\includegraphics[scale=0.75]{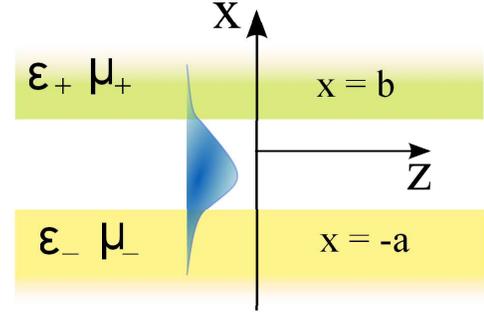}}
\vspace*{8pt}
  \caption{(Color online) 
\   The three-layer planar waveguide consists of the core layer of transverse dimension $L=a+b$ , and 
$\varepsilon_{co}(\omega), \mu_{co}(\omega)$. The thicker claddings are located at $x=b$ and $x=-a$ and made of $\varepsilon_{\pm}(\omega),  \mu_{\pm}(\omega)$. The light modes propagate along the waveguide axis $z$.}
  \label{Fig1}
\end{figure}

In this letter we present the transparent analytic solution for the dispersion relation of the confined light modes in the  nanowaveguide structure. 
Here, we have used the mathematical method of  Kacenelenbaum B. \cite{Kacenelenbaum:1953} and  developed the perturbative analysis of term $W_n$ related to the photon mass $m_o\neq 0$ manifested in the Hamiltonian function of photon in any type of waveguide regardless of the physical features of its walls.  
To the best of our knowledge this approach is accomplished here  for the first time to the analytical solution which can be applied with high accuracy to the theoretical study of the real waveguide structures.

\textit{The perturbative approach}.  
We demonstrate our approach for the well-known three layer planar waveguide consisting  of the thin slab and two much thicker claddings as depicted in Fig. \ref{Fig1}. 
This is the simplest non-trivial system where the salient features of the current approach can be  clearly demonstrated.
The eigen confined light  modes  are expressed here through the Hertz potentials \cite{Born-Wolf:1999:PO}:

\begin{equation}
\begin{array}{l}
\psi_{n}=\begin{cases}
 A_{n}e^{-k_{+}x}, &\text{$x>b$ },\\
B_{n}\sin{k_{n}^{\perp} x}+C_{n}\cos{k_{n}^{\perp} x}, &\text{$x \in (-a;b) $},\\
D_{n}e^{k_{-}x}, &\text{$x<-a$},
\end{cases}
\end{array}
\label{Equation-1_}
\end{equation}
\noindent

\noindent
where $e^{i (k_{z}z-\omega t)}$ is omitted (for notaional convenience),
$k_{z}$ is the wavenumber parallel to the interface,
$k_{n}^{\perp}$ is the transverse wavenumber in the core, $k_{z}^2+{k_{n}^{\perp 2}}=k_{0}^{2}\varepsilon_{co} \mu_{co}$, 
$k_{\pm}$ are the transverse wavenumbers in the claddings  $k_{\pm}^2=k_{z}^2-k_{0}^{2}\varepsilon_{\pm} \mu_{\pm}$ 
(where indexes $"co"$ and $"\pm"$ denote the core and the claddings, see also Fig.\ref{Fig1}). 
In this letter we restrict our attention to the transverse magnetic (TM)  modes (the transverse electric (TE)  modes can be studied in a similar way) for which the  electric and magnetic components of the modes are written as follows:

\begin{equation}
\begin{array}{l}
\vec{E}_{n}=\left( \frac{\partial^2 \psi_{n}}{\partial z \partial x},0,\frac{\partial^2 \psi_{n}}{\partial z^2}\right), 
 \vec{H}_{n}=\left(0, ik_{0}\varepsilon(x) \frac{\partial \psi_{n}}{\partial x},0\right)
\end{array}.
\label{Equation-2_}
\end{equation}

\noindent
From the boundary conditions for the fields at the two interfaces $x=-a$ and $x=b$ 
we can get the dispersion relation in the form of transcendental equation similar to \cite{Adams1981}:

\begin{equation}
\tan\left( k_{n}^{\perp}(a+b)-n\pi\right)=\frac{-\left( \frac{\varepsilon_{+}}{k_{+}}+\frac{\varepsilon_{-}}{k_{-}}\right) \frac{\varepsilon_{co}(\omega)}{k_{n}^{\perp}}}{\left( \frac{\varepsilon_{co}(\omega)}{k_{n}^{\perp}}\right)^2 -\frac{\varepsilon_{+}}{k_{+}} \frac{\varepsilon_{-}}{k_{-}}}
\label{Equation-3_}.
\end{equation}
 
\noindent

    Unfortunately, the well-known theoretical approaches permit only the approximate  solutions of the transcendental equation in  three limiting cases:  i) near cuttoff, ii) at short-wave limit  $L \omega/c\gg1$, and iii) in the strongly asymmetrical case (see, for example \cite{Yeh2008}). 
 However, the most practical systems are characterized by the intermediate sizes $L \omega/c\leq 1$ (various nanowaveguide structures etc.), by different geometries and material properties where only the numerical methods are possible for solving the complex transcendental equations \cite{Yeh2008}.  

Our aim is to develop the perturbative method \cite{Kacenelenbaum:1953} for the solution of the original wave equation $\Delta\psi_{n}+k_{n}^{\perp\,2} \psi_{n}=0$  by taking into account $W_{n}=\left(\hbar ck_{n}^{\perp}\right)^2$ and the fundamental Hamiltonian form of the dispersion relation $\omega^2=(ck_{n}^{\perp})^2+(ck_{z})^2$.
Here, the transverse wavenumber square $k_{n}^{\perp 2}$ and the eigen mode $\psi_{n}$ are decomposed in a series expansion with respect to a small parameter  $\abs{q}<1$ (the physical meaning of the parameter $q$  is determined by the parameters of the material claddings that will be discussed later):

\begin{equation}
\begin{array}{l}
k_{n}^{\perp\,2}={k_{n\, (0)}^{\perp 2}}+q {k_{n\,(1)}^{\perp 2}}+...+q^j {k^{\perp 2}_{n\,(j)}}+...,\\
\psi_{n}=\psi_{n}^{(0)}+q \psi_{n}^{(1)}+...+q^{j} \psi_{n}^{(j)}+...\\
\end{array}
\label{Equation-4_}
\end{equation}

\noindent
By inserting  Eq.\eqref{Equation-4_} in the wave equation and equating the terms of the same order to the small parameter $q$, we obtain  
$\Delta\psi_{n}^{(0)}+k_{n\, (0)}^{\perp\, 2}\psi_{n}^{(0)}=0$ and $\Delta\psi_{n}^{(1)}+k_{n\,(1)}^{\perp\, 2}\psi_{n}^{(0)}+k_{n\,(0)}^{\perp\,2}\psi_{n}^{(1)}=0$. One can (i) multiply the first equation by $\psi_{1}$, the second by $\psi_{0}$, (ii) integrate the equations over the waveguide cross section $S$ by taking into account Green's theorem and normalization $k_{n\,(0)}^{\perp 2}\int{dS\psi_{n}^{(0)\,2}}=1$,  and (iii) subsequent subtraction of two integrals leads to: 

\begin{equation}
k_{n\,(1)}^{\perp 2}=\oint_{C}{dC\left(\psi_{n}^{(1)} \dfrac{\partial \psi_{n}^{(0)}}{\partial N}-\psi_{n}^{(0)} \dfrac{\partial \psi_{n}^{(1)}}{\partial N}  \right)}_{\bigl|C},
\label{Equation-5_}
\end{equation}

\noindent
where $C$ is the contour of the cross section $S$, $\vec{N}$ is a unit vector normal to the interface between the core and claddings.

The solution $k_{n\,(1)}^{\perp 2}$  is expressed through the boundary values of the functions $\psi_{n}^{(0)}$ and $\psi_{n}^{(1)}$ on the contour $C$. 
In order to find the functiond $\psi_{0}$, $\psi_{1}$ on the $C$, we apply the Rytov-Leontovich boundary condition successfully used for studing the light effects on  the interface between two different materials \cite{Yuferev:2008:SIBc}.
This condition  determines the relation  between the magnetic and electric fields of light mode through the impedance $\zeta_{\pm}$ of claddings at the boundary contour $C$:

\begin{equation}
{\vec{E}_{co}}{\,}_{\bigl|C}\cong\zeta_{\pm}\left[\vec{N} \text{x} \vec{H}_{co}\right]_{\bigl|C},
\label{Equation-RL_}
\end{equation}

\noindent
where $\zeta^2_{\pm}(\omega)=\mu_{\pm}(\omega)/ \varepsilon_{\pm} (\omega)$ and it is assumed that $\zeta_{\pm}$  has the same order of smallness as parameter $q$.

Substituting Eqs. \eqref{Equation-2_},  \eqref{Equation-4_} in Eq.  \eqref{Equation-RL_} and equating the  zero-order terms, we find  ${\psi_{n}^{(0)}}_{\bigl|C}=0$,
this corresponds to the TM mode in the waveguide with perfect metal cladding. 
Here, we have  
 	$\psi_{n}^{(0)}=B_{n}\sin({k_{n\,(0)}^{\perp} x})+C_{n}\cos({k_{n\,(0)}^{\perp} x})$, where $B_{n}$ and $C_{n}$  	
are determined from the boundary conditions on the countour C and by normalization condition of $\psi_{n}^{(0)}$, $k_{n\,(0)}^{\perp\,2}=\left(\frac{\pi n}{b+a}\right)^2$.
By performing the same calculations with Eqs. \eqref{Equation-2_},  \eqref{Equation-4_} in Eq.  \eqref{Equation-RL_} for the first order of smallness  $\zeta_{\pm}$ and by taking into account that ${\psi_{n}^{(0)}}_{\bigl|C}=0$, we get for ${\psi_{n}^{(1)}}_{\bigl|C}$:

\begin{equation}
{q\,\psi_{n}^{(1)}}_{\bigl|C}=-\frac{ik_{0}\epsilon_{co}}{k_{0}^{\perp 2}}{\zeta_{\pm}}\frac{\partial \psi_{0}}{\partial N}_{\bigl|C}.
\label{Equation-7_}
\end{equation}

\noindent
Using the fact that the impedance is independent of the transverse coordinates and substituting  $\psi_{n}^{(0)}$, and $\psi_{n}^{(1)}$ in Eq. \eqref{Equation-5_}, then  we find  the analytical expression for the wave number in the asymmetric waveguide up to first order of the pertubation expansion:

\begin{equation}
\begin{array}{llr}
 k_{\perp}^2={\left( \frac{\pi n}{b+a}\right)}^{2}-(\zeta_{+}+\zeta_{-}) \frac{2ik_{0}\varepsilon_{d}}{b+a},\\
k_{z}^2=k_{0}^2 \varepsilon_{d}\mu_{d}-{\left( \frac{\pi n}{b+a}\right)}^{2}+(\zeta_{+}+\zeta_{-})\frac{2ik_{0}\varepsilon_{d}}{b+a}.
\end{array}
\label{Equation-8_}
\end{equation}

In case of the symmetric waveguide ($\zeta_{+}=\zeta_{-}=\zeta_{cl}$ and  $b=a$), the dispersion relation of even modes is given by:

\begin{equation}
\begin{array}{llr}
 k_{\perp}^2={\left( \frac{\pi n}{a}\right)}^{2}-\zeta_{cl} \frac{2ik_{0}\varepsilon_{d}}{a},\\
k_{z}^2=k_{0}^2 \varepsilon_{d}\mu_{d}-{\left( \frac{\pi n}{a}\right)}^{2}+\zeta_{cl} \frac{2ik_{0}\varepsilon_{d}}{a},\\
\end{array}
\label{Equation-9_}
\end{equation}

\noindent
and for the odd field modes we have:
\begin{equation}
\begin{array}{llr}
 k_{\perp}^2={\left(\frac{\pi}{2a}+ \frac{\pi n}{a}\right)}^{2}-\zeta_{cl} \frac{2ik_{0}\varepsilon_{d}}{a},\\
k_{z}^2=k_{0}^2 \varepsilon_{d}\mu_{d}-{\left(\frac{\pi}{2a}+ \frac{\pi n}{a}\right)}^{2}+\zeta_{cl} \frac{2ik_{0}\varepsilon_{d}}{a},\\
\end{array}
\label{Equation-10_}
\end{equation}
\noindent
where  $n = 0,1,2, ...$ defines a set of waveguide modes.

The series of the analytical solutions \eqref{Equation-8_}-\eqref{Equation-10_} are the main  result of this work. It is interesting to compare the solutions with the numerical results obtained for the waveguides with realistic physical parameters.

 \textit{Comparison with the numerical results}. The intensive   numerical studies of the light field propagation in the planar nano waveguide with the metamaterial claddings have been performed  in the recent works \cite{Lavoie2012}, \cite{pra-2013}.  
 This analysis required quite large computing resources. 
In  case of the metamaterial-dielectric interface, the light field can be pushed out from the metamaterial into the dielectric core when dielectric permittivity and magnetic permeability of the metamaterial are both negative. 
As a result, the low-losses field modes  are excited on the dielectric/metamaterial interface  \cite{PhysRevLett.101.263601}, \cite{PhysRevA.81.033839}.  Nanowaveguides \cite{Berini2009}  with such metamaterial claddings seem to be promising for realization of highly confined low-losses modes. 

We compare the numerical results for TM even modes \cite{Lavoie2012, pra-2013} with our analytical solution Eq. \eqref{Equation-9_}. The metamaterial claddings were described by the Drude-like model of permittivity and permeability:

\begin{equation}
\begin{array}{l}
\varepsilon(\omega)/\varepsilon_{0}=1-\omega_{e}^{2}/(\omega(\omega+i\gamma_{e})),\\
\mu(\omega)/\mu_{0}=1+F\omega^2/(\omega_{0}^{2}-\omega(\omega+i\gamma_{m})),
\end{array}
\label{Equation-11_}
\end{equation}
\noindent

\noindent
where the electric $\gamma_{e}=2.73 \cdot 10^{13} s^{-1}$ and magnetic $\gamma_{m}=\gamma_{e}$ damping rates are much less than the carrier frequency of interest; $\gamma_{e,m}\ll\omega$,  $\omega_{e}=1.37\cdot  10^{16} s^{-1}$ is a plasma frequency of the material, $\omega_{0}=0.2 \omega_{e}$ is a binding frequency, and $F=0.5$ is a geometrical factor accounts for the magnetic oscillation strength. 

\begin{figure}
\centerline{\includegraphics[scale=0.20]{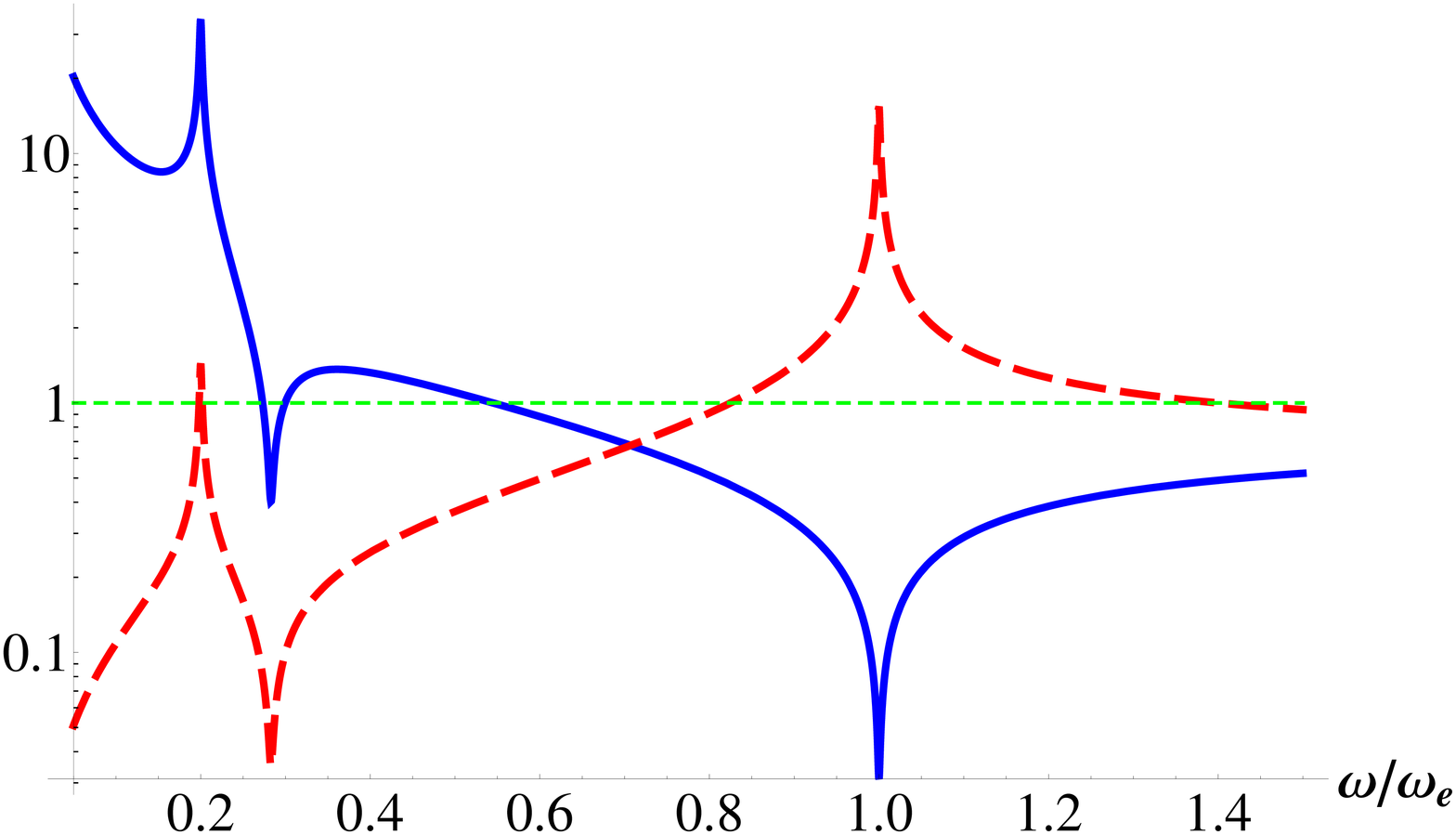}}
\caption{(Color online) Spectral behavior of the refractive index  $|n_{cl}|$ (blue line), and the impedance $|\zeta_{cl}|$  (red dashed line) of the metamaterial as functions of  $\omega/\omega_{e}$.}
\label{Fig2}
\end{figure}
\begin{figure}
\centerline{\includegraphics[scale=0.40]{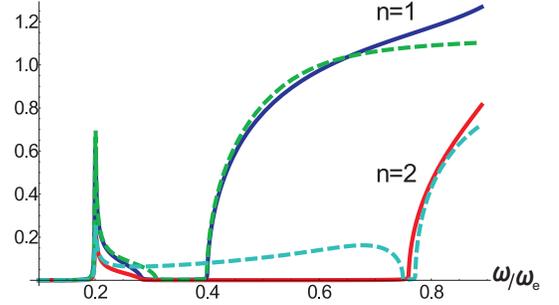}}
  \caption{(Color online) The behavior of  $Re\{k_{z}\}$ for the TM modes:  $TM_{n=1}$, $TM_{n=2}$  depending on the frequency; the analytical solutions (solid curves)  and the numerical (dashed curves) solutions are presented here for the transverse size $2a=4\pi c/\omega_{\varepsilon}\approx 275$ nm, and $\epsilon_{co}=1.3$, $\mu_{co}=1$.}
  \label{Fig3}
\end{figure}

\begin{figure}
\centerline{\includegraphics[scale=0.21]{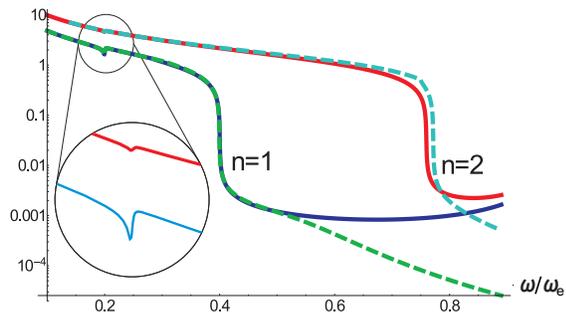}}
 \caption{(Color online) The behavior of  $Im\{k_{z}\}$ for the TM modes:  $TM_{n=1}$, $TM_{n=2}$, 
 depending on the frequency. The analytical solutions (solid curves)  and the numerical (dashed curves) solutions are given for the waveguide width $2a=4\pi c/\omega_{\varepsilon}\approx 275$ nm. The enlarged scale inset shows the sharp dips in the attenuation coefficients close to the frequency $\sim 0.2\omega/\omega_{e}$  for the both light modes (the curves demonstrate perfect coincidence between the analytical and the numerical solutions for these dips).
}
  \label{Fig4}
\end{figure}

The spectral properties of the impedance $ \zeta_{cl} (\omega)$ and the refractive index $n_{cl}(\omega)$ ($n_{cl}^2(\omega)=\varepsilon (\omega)\mu(\omega)$) 
for used metamaterial cladding are presented in Fig. \ref{Fig2}.
 As it is seen in Fig. \ref{Fig2}, the  condition  $|\zeta_{cl}| <1$  is satisfied for the frequencies  $\omega<0.85 \omega_{e}$ (except small area around $\omega =  0.2 \omega_{e}$), so the perturbative expansion Eq. \eqref{Equation-4_} is valid for most of the spectral range. 
It is seen in Figs. \ref{Fig3},  \ref{Fig4} that
each TM  mode has its spectral domain where the good match between the analytical and the numerical solutions occurs.
  In particular, for the first TM mode (n=1), we have found  the analytical solution $k_{z}^{an}$ for the longitudinal wave number coincides with the numerical one $k_{z}^{num}$ with precision $\frac{\abs{k_{z}^{an}}-\abs{k_{z}^{num}}}{\abs{k_{z}^{num}}}<10^{-2}$ for the spectral range 
  $\omega<0.25\omega_{e}$ (except for a small spectral range $\approx 0.04 \omega_{e}$ around $\omega\cong 0.2 \omega_{e}$) where the Rytov-Leontovich condition holds.
In the spectral range  $0.25 \omega_e<\omega<0.85\omega_e$, the precision of the wave number remains high  $\frac{\abs{k_{z}^{an}}-\abs{k_{z}^{num}}}{\abs{k_{z}^{num}}}<5\cdot 10^{-2}$ (i.e. in 5 times smaller, except the small area around cut-off frequency $0.4\omega_{e}$).
So as it is seen in Fig.4, the difference between the analytical and the numerical solutions for attenuation coefficient is negligible within the broad spectral range of the figure. The similar  high accuracy of the analytical description occurs for the attenuation coefficient  of the second TM-mode (n=2).

As seen in Fig. \ref{Fig4}, the small dips in the attenuation coefficient for the both TM light modes demonstrate an emergence of the low-losses field modes. 
Our preliminary analysis of the analytical solution Eq. \eqref{Equation-9_} in this spectral range shows  the spectral dip parameters (width and depth) are highly dependent on the waveguide transverse size, the field interference effects and the intensity distribution in the waveguide cross section. 
A clearer understanding of the main features  in the suppression of the low loss mode attenuation requires further detailed studies for various parameters of the waveguide.

We have also performed the accuracy analysis  of the analytical solutions for the TM mode ($n=1$). By taking into account the spectral behavior of the impedance and the refractive index we conclude the following statement.  The solution Eq. \eqref{Equation-9_} has a large accuracy within the quite broad spectral range where the impedance is sufficiently low $\abs{\zeta_{cl}}<1$ and moreover it occurs even for the   $\abs{\zeta_{cl}}\sim 1$ provided the refractive index of cladding materials is sufficiently large  $\abs{n_{cl}}>\abs{\zeta_{cl}}$. Herein, the  higher orders of the perturbation terms in series Eq. \eqref{Equation-4_} make a negligible contribution in the analytical solution.

\textit{Conclusion.} In this letter we have analytically solved the critical problem of the dispersion relation of the field modes in the  waveguide where light modes are strongly localized.  Herein, we have developed the perturbative approach for light propagation in the nanowaveguide structures that is based on the properties of Hamiltonian function of the light fields in the transversally confined waveguides.   
The perturbation approach is constructed for the decomposition of the transverse wavenumber in a small parameter $q$ determined by the physical properties of the waveguide claddings.
The obtained analytical results have demonstrated a very good accuracy compared with the well-known numerical results  within the broadband spectrum.

Thus, we have developed the theory demonstrating a unique predictive opportunity for the analytical study of the complicated waveguides characterized by small and even moderate impedance of their claddings. 
The obtained analytical solution is more convenient and practical for studying the light field modes in the waveguides, with various physical and geometrical parameters, which are important components in the conventional optics, nanooptics, integrated QED- circuits as well as in the nanoplasmonics. 
 
N.M.A. and S.A.M. thank the Russian Scientific Fund through the grant no. 14-12-01333 for financial support.

\bibliographystyle{apsrev4-1}
\bibliography{bib}

\end{document}